\documentstyle[12pt]{article}
\begin{document}
\rightline{RU-94-40}
\rightline{June 1994}
\baselineskip=18pt
\vskip 0.3in
\begin{center}
{\bf \LARGE  Note Added to \\
``Baryon Asymmetry of the Universe\\
in the Standard Model"\footnote{
CERN-TH.6734/93, RU-93-11, hep-ph/9305275
version of Nov. 22, 1993 which replaced the original version of May
17, 1993.  To be published in Phys. Rev. D.}}\\
\vspace*{0.3in}
{\large Glennys R. Farrar}\footnote{Research supported
in part by NSF-PHY-91-21039} \\
\vspace{.05in}
{\it Department of Physics and Astronomy \\ Rutgers University,
Piscataway, NJ 08855, USA}\\
\vspace{0.1in}
{\large M. E. Shaposhnikov} \\
\vspace{.05in}
{\it TH Div, CERN and INR, Moscow}\\
\end{center}
\vskip  0.1in
{\bf Abstract:}
Recent papers by Gavela et al and Huet et al claim to have shown that
incluion of higher order interactions between quasiparticles
dramatically decreases the baryon asymmetry of the universe which can
arise in the Minimal Standard Model.  These papers employ an
inconsistent calculational scheme which, for instance, violates
unitarity.  We argue that their method cannot be considered as
reliable, and thus their conclusions cannot be considered as
justified.
\thispagestyle{empty}
\newpage
\addtocounter{page}{-1}
\newpage

In our paper we estimated the baryonic asymmetry of the universe which
was produced by MSM interactions during the ew phase transition,
through CP violation in the reflection of quasiparticles from the
bubble wall.  The next corrections to our result (down by a factor
$\alpha_s$) involve processes in which reflection as well as
scattering from other particles in the plasma (which themselves may
reflect from the wall) are occuring simultaneously.  We emphasized
(section 10.4) the importance of understanding these higher order
effects before one could have confidence in the conclusion that the
MSM may account for the observed bau.  However developing a formalism
for correctly including these higher order processes is quite
non-trivial and requires construction of a kinetic theory (or, more
generally, real time Green's function approach) describing processes
near the domain wall.

Recently, papers have appeared\cite{ghop,hs} claiming that the result
of including these effects is to drastically diminish the bau produced
in the MSM.  In lieu of working in a consistent field-theoretic
formalism, these papers adopt the ad hoc procedure of modifying the
single-particle Dirac equation to include a finite lifetime for the
quasiparticle.  One manifestation of the inconsistency of this
procedure is that it violates unitarity, making physical predictions
ill-defined.  The usual kinetic equation \cite{uhlenbeck} uses matrix
elements computed from a unitary S matrix, and dissipative processes
emerge through solution of the kinetic equation rather than by any
modification of quantum mechanics.  We expect the same will be true in
the kinetic theory approach applied to this problem.

Another deficiency of the treatment of refs. \cite{ghop,hs} is that it
excludes {\em by assumption} the strong interaction phase relations
which order-by-order can compensate the loss of total reflection which
is introduced by their ad hoc inclusion of an imaginary part of the
fermionic Green's function into the Dirac equation or ad hoc
assumptions about decoherence\cite{ghop,hs}.  As long as total
reflection is the only source of the CP conserving phase which must be
present in order to produce a CP violating difference in rates,
diminishing total reflection necessarily diminishes the asymmetry.
However this need not be the case in a complete calculation,
consistently including particle interactions, since ordinary strong
interaction phase shifts can serve the function of providing a CP
conserving phase.

To see that higher order scattering from gluons need not dramatically
wash out the effect, consider more microscopically what happens when a
quasi-particle scatters from a gluon while interacting with the wall.
Recall that the CP-violation occurs because of the quantum mechanical
sum over the different paths in flavor space that the quasi-particle
can take.  We discussed in section 8 how this occurs in our
approximation, in which we included forward scattering of the quarks
with the charged W's and Higgs of the heat bath by using the one-loop
quasiparticle propagator.  It is even more transparent when one
considers the leading multi-particle process: \begin{equation} q_L^i +
\{W, H\} + wall \rightarrow q_R^j + \{W,H\} + wall.
\label{multiparticle} \end{equation} In this case, the interference
arises from the coherent sum on amplitudes for the intermediate state
to contain a quark of flavor $k$.  How does the interference between
these paths differ when there is an additional interaction with a
gluon somewhere in the process?  Some phase shift of a random nature
will be introduced into both the CP conserving and violating parts,
but it is the the same for both, and is the same for each flavor.  Thus
the only change in the result is that the particular quark under
consideration contributes with an effectively different energy than it
would have done without the gluonic scattering.  However the
contributions of different energies do not cancel, because the
reflection phase shift is always between 0 and $\pi$ as noted in
section 8, so this is not a significant effect.

What does matter and can modify the result, are processes in
which the quantum coherence in flavor space is lost due to flavor
changing interactions such as $ q + gluon \rightarrow q' + Higgs $.
The time scale associated with these interactions is $\tau_{coh} \sim
(g_s^2 f^2 T)^{-1}$ with $f$ being the Yukawa coupling constant.  This
is to be compared to the typical flavor oscillation time due to
forward scattering on the Higgs particles, $\tau_{osc} \sim (p_i -
p_j)^{-1}\sim g_s/(f^2 T)$.  Since the coherence time is
parametrically larger than the time requied for the flavor oscillation
which gives rise to the CP violating interference, i.e., $\tau_{coh}
\sim g_s^{-3} \tau_{osc}$, it is consistent not to take into account
these effects.  They make a higher order correction to the leading
result, corresponding to additional real Higgs or $W^{\pm}$'s in the
initial or final states.

In the absence of employing a legitimate, well-defined approximation
scheme, reliable results are not assured.  Thus the the claims of
refs. \cite{ghop,hs} are not justified by the work reported in them.
Generation of the observed bau by MSM physics must be considered to be
an open possibility until the necessary first-principles methods have
been developed and applied to the problem.

\end{document}